\documentclass[prd,preprintnumbers,superscriptaddress,twocolumn,floatfix,nofootinbib]{revtex4-1}
\usepackage{graphicx}

\newcommand{\bfm}[1]{\mbox{\boldmath$#1$}}

\newcommand{\gsim}{\;\rlap{\lower 3.5 pt \hbox{$\mathchar \sim$}} \raise 1pt \hbox {$>$}\;}
\newcommand{\lsim}{\;\rlap{\lower 3.5 pt \hbox{$\mathchar \sim$}} \raise 1pt \hbox {$<$}\;}

\begin{document}

\title{
\boldmath Coulomb Artifacts  and Bottomonium Hyperfine Splitting
in Lattice NRQCD.
\unboldmath}
\author{T. Liu}
\affiliation{Department of Physics, University of Alberta, Edmonton, Alberta T6G 2J1, Canada}
\author{A.A. Penin}
\affiliation{Department of Physics, University of Alberta, Edmonton, Alberta T6G 2J1, Canada}
\affiliation{Institut f\"ur Theoretische Teilchenphysik,
Karlsruhe Institute of Technology, 76128 Karlsruhe, Germany}
\author{A. Rayyan}
\affiliation{Department of Physics, University of Alberta, Edmonton, Alberta T6G 2J1, Canada}

\preprint{ALBERTA-THY-09-16}

\begin{abstract}
We study the role of the  lattice artifacts associated with the Coulomb binding
effects in the analysis of the heavy quarkonium  within lattice NRQCD. We
find that a ``na\"ive'' perturbative matching generates  spurious linear Coulomb
artifacts, which result in a large systematic error in the lattice predictions
for the heavy quarkonium spectrum. This effect is responsible, in particular, for
the discrepancy between the recent determinations of the  bottomonium hyperfine
splitting in the radiatively improved lattice NRQCD
\cite{Baker:2015xma,Dowdall:2013jqa}. We show that the correct matching
procedure which provides  full control over discretization  errors is based on
the asymptotic expansion of the lattice theory about the continuum limit,
which gives $M_{\Upsilon(1S)}-M_{\eta_b(1S)}=52.9\pm 5.5~{\rm MeV}$
\cite{Baker:2015xma}.
\end{abstract}
\maketitle
The lattice simulations within the effective theory of nonrelativistic QCD
(NRQCD) \cite{Caswell:1985ui,Bodwin:1994jh} has developed into one of the most
powerful tools for the theoretical analysis of heavy quarkonium properties
\cite{Dowdall:2011wh}. This method is entirely based on first principles, allows
for  simultaneous treatment of dynamical heavy and light quarks and gives a
systematic account of the long distance nonperturbative effects of the strong
interaction. The perturbative matching of lattice NRQCD to the full theory of
relativistic continuum  QCD is thought to be well understood. One of the most
interesting applications of the method is the analysis of the bottomonium
hyperfine splitting. The latter quantity,  defined by the mass difference
$E_{\rm hfs}=M_{\Upsilon(1S)}-M_{\eta_b(1S)}$, has been a subject of much
controversy since the first observation  of the spin-singlet $\eta_b$ state in
radiative decays of the $\Upsilon(3S)$ mesons by the BaBar collaboration
\cite{Aubert:2008ba}. The measured value of the hyperfine splitting
$71.4^{+3.5}_{-4.1}$~MeV overshot the predictions of  perturbative QCD
\cite{Kniehl:2003ap} $41\pm 14$~MeV by almost  a factor of two,  well beyond the
experimental and theoretical uncertainty bands. Such a discrepancy would
indicate a serious failure of perturbative QCD in the description of the
bottomonium ground state, in clear conflict with the general concept of the
heavy quarkonium dynamics. Further experimental  studies
\cite{Aubert:2009as,Dobbs:2012zn,Bonvicini:2009hs} were consistent with the
initial measurement,  while the Belle collaboration  reported  a significantly
lower value of the splitting  $57.9\pm 2.3$~MeV  with higher  experimental
precision \cite{Mizuk:2012pb}, see Table~\ref{tab::tab1}.  The advance of
lattice NRQCD is expected to provide an accurate model-independent  prediction
and  solve the problem on the theory side. The two most recent independent
calculations of the hyperfine splitting  which fully incorporate the one-loop
radiative corrections give   $E_{\rm hfs}=52.9\pm 5.5$~MeV \cite{Baker:2015xma}
and  $E_{\rm hfs}=60.0\pm 6.4$ \cite{Dowdall:2013jqa}.  Surprisingly,  the
difference between the central values of the results is beyond the quoted error
bars. Both calculations are based on the same lattice data and the discrepancy
exceeds what  one would expect for the perturbative approximations which are
formally of the same order in the strong coupling constant $\alpha_s$. At the
same time Refs.~\cite{Baker:2015xma,Dowdall:2013jqa} rely on different methods
of perturbative matching and the inconsistency  of the results indicates that a
careful study  of the general procedure of the  radiative improvement of lattice
NRQCD is necessary.

In this paper we study a subtle problem of the lattice NRQCD analysis of the
heavy quarkonium spectrum related to the lattice artifacts associated with
the Coulomb binding effects. We show that a widely used  direct numerical
matching procedure \cite{Hammant:2011bt,Hammant:2013sca} generates  spurious
linear Coulomb artifacts and, in particular, leads to a large systematic error
in the lattice prediction for the hyperfine splitting
\cite{Dowdall:2011wh,Dowdall:2013jqa}. The problem is related to the
all-order character of the Coulomb binding effects and is naturally solved
when the perturbative matching of lattice NRQCD  is performed through the
asymptotic expansion about the continuum limit \cite{Baker:2015xma}. We show
that after  removing the spurious contribution the result of
Ref.~\cite{Dowdall:2013jqa} is in a good agreement with \cite{Baker:2015xma}.

The paper is organized as follows. In the next section we outline the
general framework and describe different approaches to the fixed order
perturbative  matching. In Sect.~\ref{sec::3} the structure of the
Coulomb lattice artifacts is studied in detail. The result
is applied to the analysis of the  hyperfine splitting in Sect.~\ref{sec::4}.
Sect.~\ref{sec::5} is our summary and conclusion.

\section{Radiative improvement and matching in  lattice  NRQCD}
\label{sec::2}
Within the  NRQCD approach the hard modes, which require a fully relativistic
analysis, are separated from the nonrelativistic soft modes. The dynamics of
the soft modes is governed by the effective nonrelativistic action given by a
series in heavy quark velocity $v$, while the contribution of the hard modes
is encoded in the corresponding Wilson coefficients. The  nonrelativistic
action can be applied in a systematic perturbative analysis of the heavy
quarkonium spectrum \cite{Brambilla:1999xj,Kniehl:2002br,Penin:2002zv}. At
the same time the action may be used for lattice simulations of the heavy
quarkonium states, which gives  full control over nonperturbative
long-distance effects \cite{Thacker:1990bm,Lepage:1992tx}. In the latter
approach the inverse lattice spacing $a$ plays a role of the effective theory
cutoff separating the hard scale $m_q$ and the soft scale $vm_q$, where $m_q$
is the heavy quark mass.

As an example, let us consider  the  spin-dependent part of the NRQCD
Lagrangian, which is responsible for the hyperfine splitting to ${\cal O}(v^4)$.
It reads (see {\it e.g.} \cite{Pineda:1998kj,Pineda:1998kn})
\begin{equation}
{\cal L}_{\sigma}={c_F\over 2m_q}\psi^\dagger
{\bfm B}{\bfm \sigma}\psi +(\psi\to\chi_c)
+d_\sigma {C_F\alpha_s\over m_q^2}\psi^\dagger {\bfm \sigma}\psi \chi_c^\dagger
{\bfm \sigma}\chi_c,
\label{eq::NRQCD}
\end{equation}
where $\bfm B$ is the chromomagnetic field,  $C_F=(N_c^2-1)/(2N_c)$ is the
$SU(N_c)$ color group factor, $\psi$ ($\chi_c$) are the nonrelativistic Pauli
spinors of quark (antiquark)   field, and we have projected the four-quark
interaction on the color-singlet state. The coefficients $c_F=1+{\cal
O}(\alpha_s)$ and $d_\sigma={\cal O}(\alpha_s)$ parameterize the quark
anomalous chromomagnetic moment and the effective local four-quark interaction,
respectively. In the given order of the NRQCD expansion in $1/m_q$  they
depend logarithmically on the effective theory cutoff $1/a$.  This dependence
can be predicted to all orders of perturbation theory by renormalization
group methods (see {\it e.g.} \cite{Penin:2004xi,Penin:2004ay}). The radiative
improvement of the action is therefore mandatory for the correct continuum limit.

The effect discussed in this paper is characteristic for the quark-antiquark
interaction and we focus on the  Wilson coefficient $d_\sigma$ of the four-quark
operator. It vanishes in the Born approximation and is determined by matching
the one-particle irreducible quark-antiquark  scattering amplitudes in QCD   and
NRQCD. The matching becomes particulary simple when the amplitude is computed at
the quark-antiquark threshold and vanishing momentum transfer. In this case the
one-loop full QCD amplitude is
\begin{eqnarray}
 M_{\rm 1PI}^{\rm QCD} &=&
\frac{C_F\alpha_s^2}{m_q^2}\left[{C_A\over 2}\log\left({m_q\over\lambda}\right)
+\left(\ln 2-1\right)T_F\right.
\nonumber\\&+&\left.\left(1-\frac{2\pi m_q}{3\lambda}\right)C_F
      \right]\psi^\dagger {\bfm \sigma}\psi \chi_c^\dagger {\bfm \sigma}\chi_c,
\nonumber \\
\label{eq::ampqcd}
\end{eqnarray}
where  $C_A=N_c$, $T_F=1/2$, and we introduced a small auxiliary gluon mass
$\lambda$ to regulate the infrared divergence.  The power enhanced $1/\lambda$
term corresponds to the Coulomb singularity of the threshold amplitude, while the
term proportional to $T_F$ is due to the two-gluon annihilation of the
quark-antiquark pair.

\begin{table}[t]
  \begin{ruledtabular}
    \begin{tabular}{l|lc}
    \multicolumn{2}{c}{Experiment}  &\\
      \hline
      BaBar, $\Upsilon(3S)$ decays\cite{Aubert:2008ba} & $71.4^{+2.3}_{-3.1}({\rm stat})\pm 2.7({\rm syst})$&\\
      BaBar, $\Upsilon(2S)$ decays \cite{Aubert:2009as} & $66.1^{+4.9}_{-4.8}({\rm stat})\pm 2.0 ({\rm syst})$&\\
      Belle, $h_b(1P)$ decays \cite{Mizuk:2012pb} & $57.9\pm 2.3$&\\
      PDG average \cite{Beringer:1900zz} & $62.3\pm 3.2$&\\
      \hline\hline
      \multicolumn{2}{c}{Theory} & \\
      \hline
      NRQCD, NLL \cite{Kniehl:2003ap} &  $41\pm 11{\rm (th)}^{+9}_{-8}(\delta\alpha_s)$   & \\
      Lattice NRQCD ${\cal O}(v^4)$ \cite{Dowdall:2011wh}& $68\pm 9$ & \\
      Lattice NRQCD ${\cal O}(v^6)$ \cite{Dowdall:2013jqa}& $60.0\pm 6.4$  & \\
      Lattice NRQCD \cite{Baker:2015xma} & $52.9\pm 5.5$   &\\
      Lattice QCD  \cite{Davies:2013dem}& $53\pm 5$ & \\
    \end{tabular}
    \end{ruledtabular}
    \caption{\label{tab::tab1} Results of high-precision experimental and
    theoretical determinations of the bottomonium hyperfine splitting in MeV.}
\end{table}

On the other hand the  lattice NRQCD result for the one-loop amplitude
to the same order in  $1/m_q$  can be written as follows
\begin{eqnarray}
M_{\rm 1PI}^{\rm NRQCD} &=&
\frac{C_F\alpha_s^2}{m_q^2}\left[-\left(\delta
+{1\over 2}\ln\left(a\lambda \right)\right)C_A-\frac{2\pi m_q}{3\lambda}C_F
\right.
\nonumber \\
&+&\left.{d_\sigma\over\alpha_s}
\right]\psi^\dagger {\bfm \sigma}\psi
\chi_c^\dagger {\bfm \sigma}\chi_c
+{\cal O}(a),
\label{eq::ampnrqcd}
\end{eqnarray}
where the nonlogarithmic nonabelian term $\delta$ depends on a particular
realization of the lattice action.  The matching  procedure  determines the
Wilson coefficient $d_\sigma$ by equating the  effective and full theory
amplitudes, Eqs.~(\ref{eq::ampqcd},\ref{eq::ampnrqcd}), to a given order in
$\alpha_s$ and $1/m_q$. The subtlety in this procedure is related to the
treatment of the terms in the NRQCD amplitude which vanish in the continuum
limit. Below we compare two different matching prescriptions currently used
in lattice NRQCD calculations.

\subsection{Expansion about the continuum limit}
\label{sec::2.1}
This  approach  has been developed in \cite{Baker:2015xma} and relies on the
formal asymptotic expansion of the lattice loop integrals about the continuum limit
\cite{Becher:2002if} to obtain the NRQCD amplitude as a series in  $a$
order by order in the heavy quark mass expansion. To the leading order in $1/m_q$
and $a$ it gives ({\it cf.} Eqs.~(\ref{eq::ampqcd},\ref{eq::ampnrqcd}))
\begin{equation}
d_\sigma=\alpha_s\left[\left(\delta +{1\over 2}L\right)C_A
+\left(\ln 2-1\right)T_F+C_F\right],
\label{eq::ds}
\end{equation}
where  $L=\ln(a m_q)$. For  the simplest  lattice action  with
no improvement for gluonic and heavy  quark fields
the method provides the analytical result
\cite{Baker:2015xma}
\begin{equation}
\delta^{\rm naive}=-{7\over 3}+28\pi^2 b_2-256\pi^2 b_3=0.288972\ldots,
\label{eq::delnaive}
\end{equation}
where the irrational constants $b_2 = 0.02401318\ldots$, $b_3 =
0.00158857\ldots$ parameterize the lattice tadpole integrals and can be
computed with arbitrary precision. For the HPQCD action \cite{Dowdall:2011wh},
which is used in real simulations,  the nonlogarithmic coefficient has been
computed  numerically \cite{Baker:2015xma}:
\begin{equation}
\delta=0.1446(28)\,.
\label{eq::delHPQCD}
\end{equation}
Note that Eq.~(\ref{eq::ampnrqcd}) has only a logarithmic singularity in $a$ in
the formal continuum limit $a\to 0$. In higher orders of the NRQCD expansion in
$1/m_q$ the asymptotic expansion includes more singular  terms with a negative
power of $a$. Such  $1/(am_q)^n$ terms are  suppressed with respect to
Eq.~(\ref{eq::ampnrqcd}) in the  region $1/a\ll m_q$, where lattice NRQCD is
applied.

\subsection{Direct numerical matching}
\label{sec::2.2}
This approach  has been originally used for the radiative improvement of lattice
NRQCD.  Within this prescription for a given action in a given order in
$\alpha_s$  the NRQCD amplitude  is computed numerically without the expansion
in $1/m_q$ and $a$. The Wilson coefficient is then determined by the difference
between the QCD and NRQCD amplitudes in the limit $\lambda\to 0$. Since no
expansion is performed, it has a nontrivial dependence on a dimensionless
variable $am_q$ and can be written as follows
\begin{equation}
d_\sigma=\alpha_s\left[{C_A\over 2}L
+\left(\ln 2-1\right)T_F+\Delta(am_q)\right],
\label{eq::dsnum}
\end{equation}
where the logarithmic and annihilation contribution are separated and given
in an analytic form.  The function $\Delta(am_q)$ can formally be expanded
in an asymptotic series
\begin{equation}
\Delta(am_q)=\sum_{n}(am_q)^n\Delta^{(n)},
\label{eq::Delseries}
\end{equation}
where the lower summation limit is negative and depends on the approximation
used for  the NRQCD action. To determine the function $\Delta(am_q)$ we use
the numerical data of the most recent analysis \cite{Dowdall:2013jqa} based
on the ${\cal O}(v^6)$ action.\footnote{In
Refs.~\cite{Dowdall:2013jqa,Dowdall:2011wh,Hammant:2011bt,Hammant:2013sca} a
different basis of the four-quark operators is used and the  Wilson
coefficient $d_\sigma/\alpha_s$ should be identified with the linear
combination ${9\over 8}(d_1-d_2)$} In Ref.~\cite{Dowdall:2013jqa} the
numerical values of the Wilson coefficient are given for three different
values of the lattice spacing corresponding to $am_q=1.95,~2.73,~3.31$, where
the actual lattice simulations are performed. Numerical simulations
\cite{Dowdall:2013jqa,Hammant:2013sca} show that in general the terms with
negative $n$ become important for significantly lower values of the lattice
spacing corresponding to $am_q\sim 1$ and can be neglected in the region
under consideration. Indeed, the numerical data are  well approximated by a
linear function with the coefficients
\begin{equation}
\Delta^{(0)}=1.31(3) ,\qquad \Delta^{(1)}=-1.52(1),
\label{eq::Delnum}
\end{equation}
where the error bars correspond to  the linear fit of the three data points.
Note that the  result of the fit is quite  sensitive to the inclusion of the
higher order  terms,  which cannot be reliably estimated due to  lack of the
numerical data but presumably have the coefficients $\Delta^{(n)}\sim 1$.
Thus the actual uncertainty of Eq.~(\ref{eq::Delnum}) can be significantly
larger. The zero-order term of the expansion can be related to the value of
the Wilson coefficient obtained through the expansion about the continuum
limit, Eq.~(\ref{eq::ds}), as follows
\begin{equation}
\Delta^{(0)}=\delta C_A +C_F=1.767(9),
\label{eq::Del0}
\end{equation}
in a rough agreement with an estimate  Eq.~(\ref{eq::Delnum}). A
characteristic feature of the result of the numerical matching  is the linear
dependence of the Wilson coefficient on $a$, which is unusual for the lattice
simulations with the improved action. It is related to the Coulomb binding
effects in heavy quarkonium discussed in the next section.

\section{Coulomb binding effects on the lattice}
\label{sec::3}
In  perturbation theory the  Coulomb binding effects shows up through the
singular $(\alpha_s/v)^n$ terms in the contribution of the $n$-loop planar
ladder diagrams. Since in an approximately Coulomb bound state $v\sim \alpha_s$,
such terms have to be resummed to all orders. In the perturbative approach
\cite{Kniehl:2002br} this is done by constructing the perturbative expansion
about the Coulomb nonrelativistic solution rather than the free quark and antiquark.
At the same time the characteristic momentum scale of the Coulomb dynamics  is
$vm_q\ll 1/a$ and the Coulomb effects are included in the lattice NRQCD
simulations along with the nonperturbative effects of strong interactions at the
scale $\Lambda_{QCD}$. The Coulomb contribution is ultraviolet finite and
therefore its effect on the matching coefficients is suppressed  by a power of
$a$, {\it i.e.} is a lattice artifact. Below we consider the role of such
Coulomb artifacts in the calculation of  the coefficient $d_\sigma$.

\subsection{One-loop Coulomb artifacts}
\label{sec::3.1}

\begin{figure}[t]
\begin{tabular}{cc}
\includegraphics[width=4.0cm]{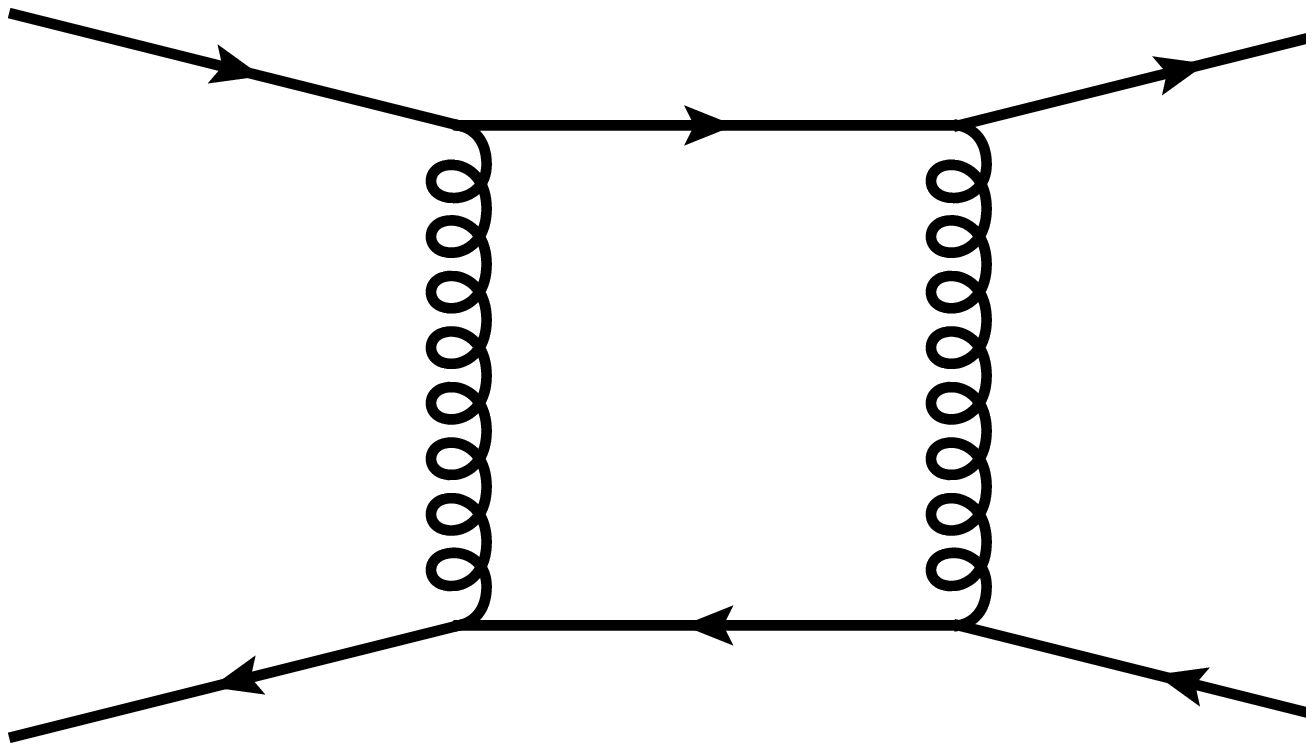}&
\hspace*{6mm}\includegraphics[width=4.0cm]{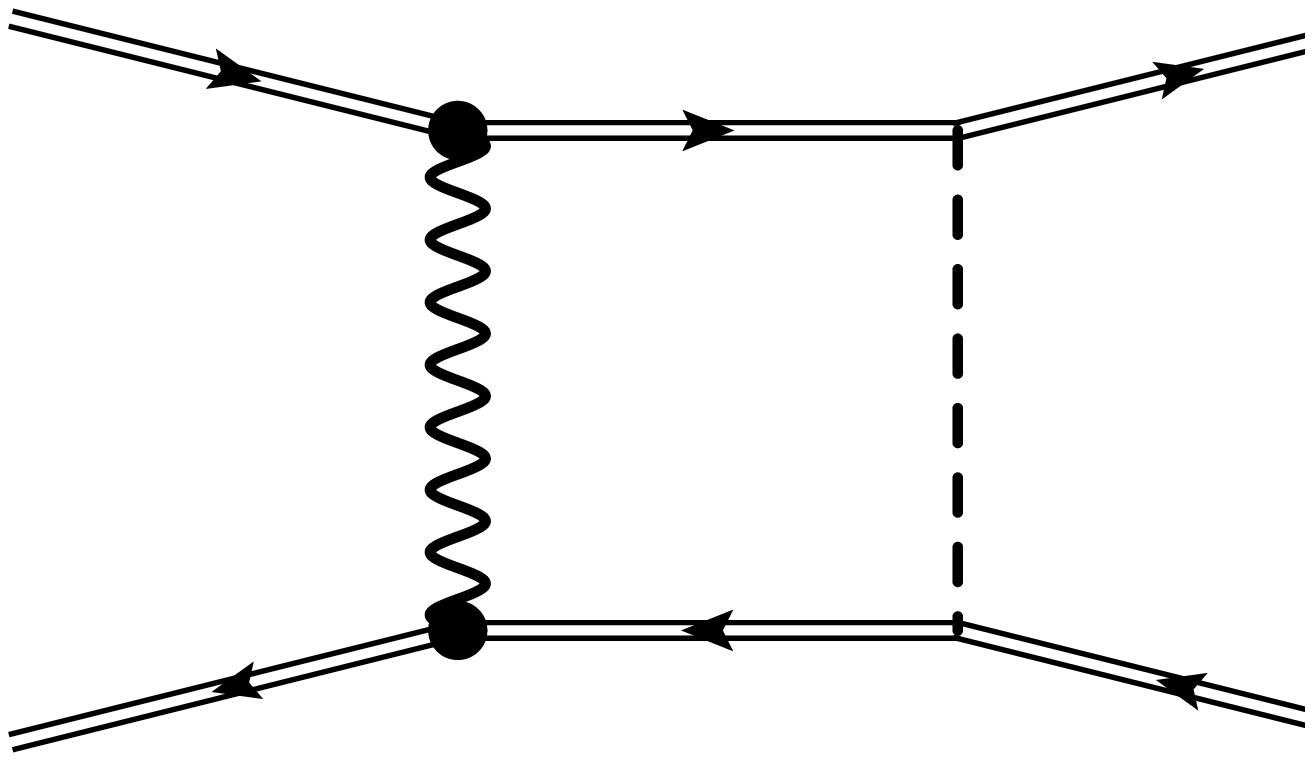}\\
(a)&\hspace*{5mm} (b) \\
\end{tabular}
\caption{\label{fig::fig1}  One-loop Feynman diagrams with Coulomb singularity
contributing to the spin-dependent one-particle irreducible part of the
scattering amplitude in   QCD (a) and NRQCD (b).  The symmetric NRQCD diagram is
not shown. In the diagram (b) the double arrow, dashed and wavy  lines stand for
the nonrelativistic quark, Coulomb and transverse gluon propagators,
respectively. The black circles denote the effective spin chromomagnetic
interaction proportional to the Wilson coefficient $c_F$ in
Eq.~(\ref{eq::NRQCD}).}
\end{figure}

The Coulomb singularity is contained in the planar box diagrams of QCD
(Fig.~\ref{fig::fig1}a) and NRQCD (Fig.~\ref{fig::fig1}b), and  takes the form
$\alpha_sm_q/\lambda$ since the matching calculation is performed with $v=0$.
Let us consider the evaluation of the corresponding  contribution to the  NRQCD
amplitude to ${\cal O}(a)$. The expansion of the lattice NRQCD Feynman rules in
$a$ generates the second or higher order terms  so we can use the continuum
expressions for the gluon and nonrelativistic heavy quark propagators
\begin{equation}
D_{\mu\nu}(k)={g_{\mu\nu}\over k^2-\lambda^2}, \qquad S(k)=
{ 1\over k_0-{{\bfm k}^2/ (2m_q)}},
\label{eq::propagators}
\end{equation}
where $k=(k_0,{\bfm k})$.  After integrating over the time component of the
virtual momentum by taking the residue of the heavy quark propagator, the Coulomb
contribution to the scattering amplitudes takes the form
\begin{eqnarray}
M_{\rm C}^{\rm NRQCD} &=&-
{2\over 3\pi}\frac{C_F^2\alpha_s^2}{m_q}\left[
\int_{\cal B}{{\rm d}{\bfm k}  \over  ({\bfm k}^2+\lambda^2)^2}
\right]\psi^\dagger {\bfm \sigma}\psi
\chi_c^\dagger {\bfm \sigma}\chi_c
\nonumber\\
&&+{\cal O}(a^2),
\label{eq::Coulomblat}
\end{eqnarray}
where the integration over the spatial virtual momentum is restricted to the
first Brillouin zone. Without loss of generality we consider a spherically
symmetric lattice with the Brillouin zone defined by $ |{\bfm k}|<\pi/a$, and
after integrating over the angular components obtain
\begin{eqnarray}
\int_{\cal B}{{\rm d}{\bfm k}  \over  ({\bfm k}^2+\lambda^2)^2} &=&
\int_0^{\pi/a}{\rm d}{|\bfm k|} { 4\pi {\bfm k}^2 \over  ({\bfm k}^2+\lambda^2)^2}
\nonumber\\
&=&
{\pi^2\over\lambda}-{4}a+{\cal O}(a^2).
\label{eq::Coulomblatint1}
\end{eqnarray}
The contribution of the first singular term of Eq.~(\ref{eq::Coulomblatint1})
agrees with Eq.~(\ref{eq::ampnrqcd}), while the second term represents the
linear Coulomb lattice artifact corresponding to $\Delta^{(1)}=-{8\over
3}{C_F\over \pi}$ in the expansion  Eq.~(\ref{eq::Delseries}). This
coefficient  is  independent of the infrared cutoff but  does depend on the
approximation for the NRQCD action.  For example, let us consider the  ${\cal
O}(v^4)$ heavy quark propagator
\begin{eqnarray}
\label{eq::quarkprop}
S(k)&=&{1\over k_0-{{\bfm k}^2/ (2m_q)}
+{{\bfm k}^4/(8m_q^3)}}.
\label{eq::propv4}
\end{eqnarray}
The correction term in  the denominator of Eq.~(\ref{eq::propv4}) results in
an additional contribution  to the integral  in  Eq.~(\ref{eq::Coulomblat})
\begin{eqnarray}
-\int_0^{\pi/a}{\rm d}{|\bfm k|}
{\pi \over (m_q^2-{\bfm k}^2/4)}
&=&
-{4}a+{\cal O}(1/m_q),
\label{eq::Coulomblatint2}
\end{eqnarray}
where we neglected the gluon mass since the integral is infrared finite. Thus
the ${\cal O}(v^4)$ correction to the  nonrelativistic kinetic energy  increases
the coefficient of the linear term  by factor two, which gives
\begin{equation}
\Delta^{(1)}=-{16\over 3}{C_F\over \pi}.
\label{eq::Del1an}
\end{equation}
For comparison with the direct numerical matching this value  should be
multiplied by a geometrical factor $\nu =0.831\ldots$, which converts the
result obtained on  the spherically symmetric lattice  into the one for the
standard cubic lattice \cite{Baker:2015xma}. This gives $\Delta^{(1)}\approx
-1.87$, which   is slightly above the ${\cal O}(v^6)$ value of
Eq.~(\ref{eq::Delnum}),  but is in a very good agreement with the value
$\Delta^{(1)}\approx -1.82$ obtained from the fit of the ${\cal O}(v^4)$
result~\cite{Hammant:2011bt}. Numerically the one-loop linear artifact
dominates the series in Eq.~(\ref{eq::Delseries}) for typical values of $a$
and one may argue that its  inclusion into the Wilson coefficient  is
mandatory. However  the above analysis takes into account only a single
Coulomb gluon exchange while the effect of multiple Coulomb exchanges is not
parametrically suppressed and significantly changes the structure of the
expansion in $a$ as discussed in the next section.

\subsection{Coulomb artifacts to all orders}
\label{sec::3.2}

Though we consider the properties of  the  heavy quarkonium  bound states, the
analysis of the previous sections involved the  scattering amplitudes of the
free quark and antiquark. This is sufficient if in the matching region the
binding effects can be expanded in a regular series in $\alpha_s$.  The Coulomb
artifacts, however, are related to the dependence of the  bound state
characteristics on the lattice spacing, which cannot be described  within the
finite-order perturbation theory. Indeed, by using the Coulomb equations of
motion the diagram in Fig.~\ref{fig::fig1}b can be absorbed into the Coulomb
wave function of an external state. Thus in this case the matching  procedure
should be applied to the matrix elements of the effective action operators
between the quarkonium states with the wave functions computed on the lattice
and in the continuum. The relevant  nonrelativistic Coulomb wave function in the
continuum is well known.  On the lattice it can be obtained  in a
straightforward way by solving the nonrelativistic Schr\"odinger equation  as a
difference equation for a given finite $a$. In the formal limit
$\Lambda_{QCD}\ll v^2m_q$ one can neglect  the nonperturbative dynamics of
strong interactions at long distance and the result obtained by numerical
solution of the discretized Schr\"odinger equation provides the same bound state
wave function as the real lattice simulations based on the functional integral
approach.

Let us apply the above ``Schr\"odinger matching'' approach to  the analysis
of the hyperfine splitting. The relevant four-quark  operator is generated by
the magnetic  gluon exchange and corresponds to the leading order
spin-dependent amplitude\footnote{In a Coulomb system the infrared
divergences are regulated by the dynamically  generated binding energy and we
can neglect the fictitious mass in gluon propagator.}
\begin{eqnarray}
M_{\rm LO}^{\rm NRQCD} &=&
-{2\over 3}\frac{C_F\alpha_s}{m_q^2}
\psi^\dagger {\bfm \sigma}\psi
\chi_c^\dagger {\bfm \sigma}\chi_c\,.
\label{eq::loamp}
\end{eqnarray}
In coordinate space this  local spin-flip operator is proportional to
$\delta({\bfm x})$. The corresponding matrix element, which in fact determines
the leading order hyperfine splitting, is proportional to $|\psi(0)|^2$, where
$\psi({\bfm x})$ is the ground state quarkonium wave function. The Coulomb
solution for this quantity takes into account the contribution of all-order
Coulomb exchange diagrams including Fig.~\ref{fig::fig1}b.  In the continuum it
reads $|\psi(0)|^2={C_F^3\alpha_s^3m_q^3/(8\pi)}$. The lattice value of the wave
function at the origin is obtained by numerical solution of the Schr\"odinger
equation with the Coulomb Hamiltonian. It is performed on a spherically
symmetric lattice, which retains the qualitative properties of the solution. To
match the setup of real lattice simulations \cite{Lepage:1992tx} we use the
central difference discretization of the kinetic energy   operator, which has
${\cal O}(a^4)$ local error. The boundary condition of the  eigenstate problem
is  determined by the value of the exact continuum solution at sufficiently large
distance, where the wave function is exponentially suppressed. Though the
parameters  of the bound state can be obtained for an arbitrary value of
lattice spacing, we are interested  in their behavior at small $a$. For the
expansion of the ground state energy and the wave function at the origin about
their continuum values we get
\begin{eqnarray}
E&=&-{C_F^2\alpha_s^2m_q\over 4}\left(1-{1\over 4}\bar{a}^2
+{\cal O }(\bar{a}^4)\right),
\label{eq::en} \\
|\psi(0)|^2&=&{C_F^3\alpha_s^3m_q^3\over 8\pi}\left(1-{1\over 2}\bar{a}^2
+{\cal O }(\bar{a}^4)\right),
\label{eq::psi}
\end{eqnarray}
where  $\bar{a}=C_F\alpha_sam_q/2$  is the dimensionless lattice  spacing in
Coulomb units, and the rational coefficients of the expansion are conjectured
from the high accuracy numerical result.  The expression for the ground state
energy is not required for our analysis and is given for completeness.
Eq.~(\ref{eq::psi}) does not have a linear dependence on $a$. This may be
expected since the integration of a second order difference equation with ${\cal
O}(a^4)$ local discretization  error gives  ${\cal O}(a^2)$ global error of the
solution (see {\it e.g.} \cite{Hairer:2008}).

Eq.~(\ref{eq::psi}) determines the difference between the lattice and continuum
results for the matrix element of the leading order spin-flip operator
Eq.~(\ref{eq::loamp}). As we see, the linear dependence of the bare result on
the lattice spacing is absent. Thus, the one-loop linear term in the Wilson
coefficient~(\ref{eq::dsnum}) in fact {\it introduces} a linear dependence  of
the  radiatively improved result on $a$ and one has to add an additional
``matching'' correction in order to compensate this dependence.
Strictly speaking the correction to the long-distance matrix element which
depends on the properties of a specific bound  state should not  be associated
with a universal NRQCD coupling and  should be consider separately.
However, the absence of the linear dependence of the bound state parameters on
the lattice spacing is a general property of the central difference
discretization and  one can account for this fact simply by setting
\begin{equation}
\Delta^{(1)}=0 
\label{eq::DelCoulomb}
\end{equation}
in the case under consideration. Thus when the Coulomb effects are taken into
account consistently to all orders in $\alpha_s$, the linear artifact in the
four-quark matching coefficient is effectively  {\it absent} and the first
nonvanishing term is quadratic in $a$.

We would like to emphasize that though the coefficient in
Eq.~(\ref{eq::psi}) is proportional to $\alpha_s^2$, it gets contributions from
all-order Coulomb exchange diagrams. This coefficient is changed by the higher
order terms in the NRQCD action and is different for the  standard cubic
lattice,  as in the case of the linear artifact discussed in the previous
section. In principle, within the same method the Coulomb lattice artifacts can
be evaluated for a given NRQCD action on a given lattice. However for practical
applications they can be removed along with the nonperturbative artifacts  by
the extrapolation of the lattice data to $a=0$,  as it is  discussed  in the
next section. The absence of the linear artifact  is crucial for this procedure
though.

Note that the $\Upsilon$ spectrum has been studied within the  discretized
Schr\"odinger-Pauli equation framework similar to the one used in this paper
but on a  more realistic lattice \cite{Bali:1998pi}. The numerical result of
Ref.~\cite{Bali:1998pi} with a good precision rules out the linear Coulomb
artifacts in the bare lattice data for the energy levels in full agreement
with our analysis.

Let us now discuss the reason of the qualitative difference between the
one-loop and all-order dependence of the  bound state parameters on $a$. As
it has been pointed out, in the one-loop calculation the leading ${\cal
O}(a)$ correction to the continuum result is due to the effective momentum
cutoff at the scale  $1/a$ while the corrections to the free continuum quark
and gluon propagators contribute only at ${\cal O}(a^2)$. For the bound quark
propagator, however,  the corrections start at ${\cal O}(a)$ due to the
Coulomb singularity and, according to Eq.~(\ref{eq::psi}), cancel the linear
term originating from the momentum cutoff. Note that this cancellation  is
specific for the lattice regularization in use. From the above analysis it is
clear that if the effective theory is regularized by a momentum cutoff
$\Lambda_{UV}\sim 1/a$ only, the linear artifacts of the form
$m_q/\Lambda_{UV}$ are indeed generated and  for a finite cutoff should  be
cancelled  by the corresponding  term in the Wilson coefficient (see, {\it
e.g.} \cite{Hill:2000qi}).

\section{Determination of the energy spectrum from the lattice data}
\label{sec::4}

\begin{figure}[t]
\begin{tabular}{c}
\includegraphics[width=9.0cm]{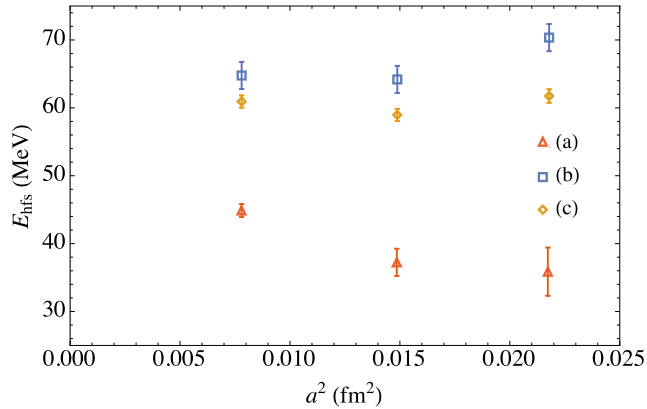}\\
\end{tabular}
\caption{\label{fig::fig2}  The results of the lattice simulation of the
bottomonium hyperfine splitting with ${\cal O}(v^6)$ NRQCD action and the
four-quark Wilson coefficient given by (a) the asymptotic expansion about the
continuum limit \cite{Baker:2015xma}, (b) the direct numerical matching and
(c)  $d_\sigma=0$ \cite{Dowdall:2013jqa}.   All  data points include the
statistical error and the uncertainty in the value of the lattice spacing.
The error bars of (a) include also the uncertainty due to the higher order
perturbative corrections. The difference between (a) and (b) data sets
is mainly due to the spurious  linear Coulomb artifact contributing to (b).}
\end{figure}

Let us now consider how the Coulomb artifacts affect the determination of the
energy spectrum from the lattice data.  The results of  nonperturbative
lattice NRQCD simulations are typically given for  $a\sim 1/(vm_b)$
\cite{Dowdall:2011wh,Dowdall:2013jqa}. The use of relatively  large values of
the  lattice spacing ensures the suppression of the unphysical $1/(am_b)^n$
contributions, which become important at $a\sim 1/m_b$ . At the same time it
results in sizable Coulomb lattice  artifacts proportional to a power of
$\alpha_s am_b\sim 1$.  In addition the  lattice data include the
nonperturbative lattice artifacts which scale as $(a\Lambda_{QCD})^2$ and
cannot be removed  through  the  matching procedure discussed above. To
minimize these effects the results of the lattice simulations are numerically
extrapolated to $a=0$. The extrapolation below  $a\sim 1/m_b$ in this case is
justified because  the numerical effect of the $1/(am_b)^n$ terms on the data
points is small. Since the radiatively improved lattice result is supposed to
be free of linear artifacts, the extrapolation is performed through a
constrained fit of the data points  by a polynomial in $a$ with  {\it
vanishing} linear term (see {\it e.g.}
\cite{Baker:2015xma,Dowdall:2011wh,Dowdall:2013jqa}).  The  correct treatment
of the linear artifacts is therefore crucial for the extrapolation procedure.
As it has been shown in the previous section  by the analysis of the discretized
Schr\"odinger equation, the linear Coulomb artifacts are absent in the bare
lattice data.  The contribution of  the four-quark interaction to $E_{\rm
hfs}$ reads
\begin{equation}
\Delta E_{\rm hfs}=-d_\sigma{4 C_F\alpha_s\over m_q^2}|\psi(0)|^2.
\label{eq::DelE}
\end{equation}
Thus the  linear Coulomb artifact  in the Wilson coefficient obtained by the
direct numerical matching \cite{Hammant:2011bt, Hammant:2013sca} results  in
spurious linear dependence of the radiatively improved lattice data on $a$,
which leads to a systematic error in the extrapolation procedure based on the
fit with the vanishing linear term. At the same time the Wilson coefficient
obtained by the asymptotic expansion about the continuum limit is free of the
Coulomb artifacts and provides the correct functional dependence of the
radiatively improved lattice data on $a$ and therefore can be used for
consistent extrapolation procedure. The numerical effect of the spurious
linear artifact turns out to be  very significant. In Fig.~\ref{fig::fig2} we
compare the ${\cal O}(v^6)$ lattice NRQCD result for  the bottomonium
hyperfine splitting with the four-quark Wilson coefficient obtained  by the
asymptotic expansion about the continuum limit \cite{Baker:2015xma} and
through the direct numerical matching \cite{Dowdall:2013jqa}. As a reference
point  we also present the numerical data for $d_{\sigma}=0$. The difference between
the results obtained within the two matching schemes  is mainly due to the
contribution of the linear artifact. It can be as large as a hundred percent
for the actual values of lattice spacing and remains significant after the
extrapolation  to $a=0$ is performed. The analysis \cite{Baker:2015xma} with
the matching coefficient Eq.~(\ref{eq::ds}) after the extrapolation  gives
$E_{\rm hfs}=51.5\pm 5.7$~MeV. At the same time the analysis
\cite{Dowdall:2013jqa} gives $E_{\rm hfs}=60.0\pm 6.4$~MeV.  The discrepancy
between the central values is well beyond  the reported discretization/extrapolation
uncertainty, which is below $3$~MeV. Thus the analysis  of the hyperfine
splitting in
Refs.~\cite{Hammant:2011bt,Hammant:2013sca,Dowdall:2011wh,Dowdall:2013jqa}
contains a systematic error and should be corrected.

The result of the direct  numerical matching  can yet be used for the
self-consistent analysis of the quarkonium spectrum through the decomposition of
the form of Eqs.~(\ref{eq::dsnum},\ref{eq::Delseries}). After separating the
logarithmic part, the result for the Wilson coefficient should be fitted by a
polynomial in $am_q$ and the linear term of the expansion should be subtracted.
In the case under consideration only the $\Delta^{(0)}$ term should be retained
in $d_\sigma$. The further analysis  follows Ref.~\cite{Baker:2015xma} with the
coefficient $\Delta^{(0)}$ from Eq.~(\ref{eq::Del0}) substituted by  the one
from Eq.~(\ref{eq::Delnum}). This gives the central value  $E_{\rm
hfs}=52.7$~MeV, which is outside the error interval of Ref.~\cite{Dowdall:2013jqa}
but in a very good agreement with the ${\cal O}(v^6)$ result of
Ref.~\cite{Baker:2015xma} given above.

Though the quadratic  Coulomb artifact is eliminated by extrapolation, it
is instructive  to estimate its contribution to the dependence of the
lattice data on $a$ and corresponding uncertainty in the the extracted value
of $E_{\rm hfs}$.  The result  of the  fit for the hyperfine splitting can be
represented as follows
\begin{equation}
E^{\rm lattice}_{\rm hfs}=E_{\rm hfs}\left(1-(\Lambda a)^2+{\cal O}(a^3)\right),
\label{eq::fit}
\end{equation}
where $\Lambda$ is the mass  scale characterizing  the approach of the lattice
approximation to the continuum limit. Numerically one gets $\Lambda\approx
360$~MeV for the ${\cal O}(v^4)$ and $\Lambda\approx 790$~MeV for the ${\cal
O}(v^6)$ lattice action \cite{Baker:2015xma}.  On the other hand the quadratic
Coulomb artifact with the coefficient Eq.~(\ref{eq::psi})  corresponds to
\begin{equation}
\Lambda={C_F\alpha_sm_q\over 2\sqrt{2}},
\label{eq::Lam}
\end{equation}
which  gives $\Lambda\approx 530$~MeV for the values of the input parameters
taken in the middle of a typical interval  for the  lattice spacing. 

Though
Eq.~(\ref{eq::Lam}) is obtained in a simplified model  with  the Coulomb
Hamiltonian and on a spherical lattice, we can conclude that the quadratic
Coulomb artifact to a large  extent determines the dependence of the bare
lattice result on  $a$ and can be used as a prior for the constrained fit. As we
observed in Sect.~\ref{sec::3.1} the effect of the lattice artifacts is enhanced
by the relativistic corrections  since the contribution of the higher dimension
operators is more sensitive to the ultraviolet momentum region. This explains a
slower approach to the  continuum limit and larger discretization errors of the
extrapolation based on  ${\cal O}(v^6)$  lattice data. The smaller
discretization uncertainty  balances the larger relativistic corrections in the
${\cal O}(v^4)$  case and both actions provide comparable total errors. The best
estimate  is obtained as the weighted average of two results \cite{Baker:2015xma}
\begin{equation}
E_{\rm hfs}=52.9\pm 5.5~{\rm MeV},
\label{eq::fin}
\end{equation}
which is  $1.4$~MeV  above the  ${\cal O}(v^6)$ value with slightly reduced
error. Hence Eq.~(\ref{eq::fin}) can be considered as an unambiguous and
the most accurate lattice NRQCD prediction for the bottomonium hyperfine
splitting available so far. It is interesting to compare this result to the
most recent analysis of the bottomonium hyperfine splitting within  lattice
QCD \cite{Davies:2013dem}. A fully relativistic description of the bottom
quark is still beyond the reach of the lattice simulations due to the large
value of $m_b$ compared to typical hadronic scale. In
Ref.~\cite{Davies:2013dem} the result for the bottomonium system is obtained
by extrapolating the fictitious lighter quarkonium spectrum to the physical
value of the bottom quark mass.  Such an extrapolation gives $E_{\rm
hfs}=53\pm 5$~{\rm MeV}, in a very good agreement with the NRQCD result
Eq.~(\ref{eq::fin}).

\section{Summary and conclusion}
\label{sec::5}
In this paper we critically examined the matching procedure for the radiative
improvement of the lattice  NRQCD. We have demonstrated that the Wilson
coefficients of the effective four-quark interaction obtained by the widely
used direct numerical matching suffer from  spurious linear Coulomb lattice
artifacts, which result in a large systematic error  in the predictions for
the heavy quarkonium spectrum.  This problem is solved by using the matching
procedure based on the asymptotic expansion about the continuum limit. We
also have shown how the direct numerical matching should be modified for a
consistent treatment of the lattice artifacts.

Our analysis resolves  the discrepancy between the most recent lattice NRQCD
predictions for the bottomonium hyperfine splitting
\cite{Baker:2015xma,Dowdall:2013jqa}  in favour of the result
of Ref.~\cite{Baker:2015xma}, Eq.~(\ref{eq::fin}).

\vspace{5mm}

\acknowledgements
We would like to thank Marc Baker, Antonio Pineda and Nikolai Zerf for useful
discussions and collaboration.  This work  is supported in part by NSERC.
The work of A.P. is supported in part by the  Perimeter Institute for Theoretical Physics.


\end{document}